\documentclass[12pt]{article}
\usepackage{amsmath,amsthm,amsfonts,amssymb}
\usepackage{graphicx}
\usepackage{enumerate}
\usepackage[authoryear]{natbib}
\usepackage{url} 
\usepackage{array}
\usepackage{amscd}
\usepackage{color}
\usepackage{mathrsfs}
\usepackage{dcolumn}
\usepackage{latexsym}
\usepackage{bm,bbm}
\usepackage{extarrows}
\usepackage{bm,bbm}
\usepackage{rotating}
\usepackage{multirow,bigstrut,booktabs}
\usepackage[colorlinks,citecolor=blue,urlcolor=blue,linkcolor=blue]{hyperref}
\usepackage[title]{appendix}
\usepackage{placeins}

\newtheorem{theorem}{Theorem}

\newtheorem{assumption}{Assumption}
\numberwithin{remark}{section}

\newcommand{\blind}{1}

\begin{document}

\def\spacingset#1{\renewcommand{\baselinestretch}%
{#1}\small\normalsize} \spacingset{1}


\if1\blind
{
  \title{\bf Modelling time series of counts with hysteresis}
  \author{Xintong Ma,~Dong Li\thanks{
    Li's work is supported by Beijing Natural Science Foundation (No.F251002).}\hspace{.2cm}\\
    Department of Statistics and Data Science, Tsinghua University, \\Beijing 100084, China.\\
    and\\
    Howell Tong \\
    Department of Statistics, London School of Economics,  London, U.K.}
  \maketitle
} \fi

\if0\blind
{
  \bigskip
  \bigskip
  \bigskip
  \begin{center}
    {\LARGE\bf Modelling time series of counts with hysteresis}
\end{center}
  \medskip
} \fi

\bigskip
\begin{abstract}
In this article, we propose a novel model for time series of counts called the hysteretic Poisson autoregressive (HPART) model with thresholds by extending the linear Poisson autoregressive model into a nonlinear model. Unlike other approaches that bear the adjective ``hysteretic", our model incorporates a scientifically relevant controlling factor that produces genuine hysteresis. Further, we re-analyse the buffered Poisson autoregressive (BPART) model with thresholds. Although the two models share  the convenient piecewise linear structure, the HPART model probes deeper into 
the intricate dynamics that governs regime switching.  We study the maximum likelihood estimation of the parameters of both models and their asymptotic properties in a unified manner, establish tests of separate families of hypotheses for the non-nested case involving a BPART model and a HPART model, and demonstrate the finite-sample efficacy of parameter estimation and tests with Monte Carlo simulation. We showcase advantages of the HPART model with two real time series, including plausible interpretations and improved out-of-sample predictions.
\end{abstract}

\noindent%
\vfill

\newpage
\spacingset{1.9} 
\section{Introduction}\label{sec:intro}
\subsection{Background}
Let us start by stating that our use of the term {\it hysteresis} is in line with 
the phenomenon defined in physics, engineering, biology and many other scientific and engineering disciplines. It is associated with a hysteresis curve that is a \textit{multivalued}  function with two branches. Which branch a trajectory follows depends critically on a \textit{controlling factor}. For example, the action of loading/unloading a weight to/from a spring is a controlling factor: unloading a weight returns the spring to (possibly) its original state but along a different branch than the one initially traversed by the loading of a weight on the same spring. Interestingly, as early as in 1980, \cite{TK} (p. 280) recognised the importance of hysteresis in the above sense; specifically, they referred to the cusp catastrophe that generates hysteresis among other features ({\it op cit.}, p. 292 that cited \cite{Zeeman1977}). In fact, they even fitted model (A1) ({\it op cit.}, p. 283) to the Canadian lynx data. Model (A1) uses the sign of the first difference of the observations as the controlling factor. As far as we know, model (A1)  is the first hysteretic autoregressive model for real-valued time series in the literature. 

Later,  for {\it real-valued (i.e. non-count)} time series data, \cite{Zhu:Yu:Li:2014} and \cite{G}
studied another interesting mechanism
by introducing a buffer zone.
Further development of this approach followed, e.g. 
\cite{Lo:Li:2016}, \cite{Zhu:BufferedGARCH}, \cite{Li:Zeng:2020},  \cite{Liu:Zhu:2020}, \cite{Wang:Li:2020}, and others.

Time series of counts is an important subclass of time series,
ubiquitous in the real world.
The statistical analysis of such time series has received much attention
in recent years; see, e.g., \cite{ArFokianos:2023, ArFokianos:2024}, \cite{SeasonalCTS}, \cite{GaussCountTS},
 \cite{WeissZhu2024}, and the references therein. For
comprehensive reviews, see
\cite{Dag:2012}, \cite{DHLR}, \cite{CHW}, \cite{RD}, \cite{FFKV},  \cite{Karlis2023},  \cite{Liu:Zhu:2023}, and \cite{Sellers}.

A well-known  model for time series of counts is the Poisson autoregressive (PAR) model,
usually attributed to 
\cite{FRT}; see also \cite{FLO} and \cite{SCS2007} for slightly earlier and related developments.
It has been
 further  studied extensively by
\cite{FDag:2011, FDag:2012}, \cite{Neumann}, \cite{CW}, \cite{CF:2015}, \cite{Ali:2016}, \cite{Davis:Liu:2016},   \cite{Douc:2017},  
  \cite{ModelSlection}, \cite{FokianosST:2020}, \cite{Diop2021}, \cite{Doukhan:2021, DLN:2022},  \cite{Zhu2022}, \cite{Huang:2023}, and others.

Although popular, the PAR model is generally incapable of capturing negative serial dependence among observations. Adopting the threshold approach of \cite{T}, \cite{CW} proposed a new and flexible self-excited threshold Poisson autoregressive (SETPAR) model.

Subsequently, for count data, \cite{Truong2017} introduced the buffer mechanism into the PAR model and proposed a new integer-valued time series model with the adjective {\it hysteretic}, although its use  differs from the sense  explained at the opening paragraph. Therefore, to distinguish this model from the one that we are about to develop,
we assign their buffered Poisson autoregressive model with thresholds the acronym BPART. 
 Later in this paper, we propose a hysteretic Poisson autoregressive (HPART) model, in which a hysteretic regime is introduced into the classical Poisson autoregression. We also explore the BPART model further. 

On the one hand, the BPART model and the HPART model are similar: besides being piecewise linear, they 
initially divide the state space into three regimes: an upper regime, a lower regime and an `intermediary' hysteretic/buffer regime before ending up with just the first two regimes.  On the other hand, the regime-switching mechanisms of the Poisson intensity process for data within the hysteretic/buffer regime are quite different as we shall see in Subsection \ref{twononlinear}. Briefly, in the HPART model, the mechanism involves an explicitly defined controlling factor that is based on the first difference of the two immediately past observations as the control factor, as in Model (A1) mentioned above.\footnote{The controlling factor can be a much more general function, including, e.g., a function of a covariate time series of either count data or non-count data.}  On the other hand, the BPART model does not involve such a controlling factor.


To distinguish between the two mechanisms, we develop tests for non-nested models to detect departure from a BPART/HPART model in the direction of a HPART/BPART model, in the context of separate families of hypotheses initiated by \cite{Cox1960,Cox1962}. See also \cite{Cox2013}.

\subsection{Two nonlinear mechanisms}\label{twononlinear}
Generally, a nonnegative integer-valued time series $\{y_t: t\in\mathbb{Z}\}$ is said to follow a parametrically
stochastic dynamic Poisson  model, if $y_t$ conditionally follows a Poisson distribution, namely
\begin{flalign*}
\mathcal{L}(y_t\mid\mathcal{F}_{t-1})=\mathrm{Poisson}(\lambda_t),\quad t\in\mathbb{Z},
\end{flalign*}
where $\mathcal{F}_{t-1}$ is a sequence of $\sigma$-fields generated by $(y_j: j\leq t-1)$, and the intensity process $\lambda_t=g_{\theta}(y_{t-1},\lambda_{t-1},...)$ is a positive function depending on parameter $\theta$ and is measurable with respect to $\mathcal{F}_{t-1}$ for all $t\in\mathbb{Z}$, i.e., $\lambda_t\in \mathcal{F}_{t-1}$.

(a) A count time series $\{y_t: t\in\mathbb{Z}\}$ is said to follow a first-order BPART model,
if the intensity process $\lambda_t$ satisfies the stochastic recurrence
\begin{flalign}\label{bpart}
\lambda_t =
	\begin{cases}
		\omega_1 + \alpha_1y_{t-1}+\beta_1\lambda_{t-1}, & \mathrm{if} \hspace{2mm} R_{t}=1,\\\\
		\omega_2  + \alpha_2y_{t-1}+\beta_2\lambda_{t-1}, & \mathrm{if} \hspace{2mm} R_{t}=0,
	\end{cases}\qquad
R_{t} =
	\begin{cases}
		1, & \mathrm{if} \hspace{2mm} y_{t-1}\leq r,\\
		0, & \mathrm{if} \hspace{2mm}  y_{t-1}> s,\\
		R_{t-1}, & \mathrm{otherwise},
	\end{cases}
\end{flalign}
where 
$\omega_i>0,\alpha_i\geq0,\beta_i\geq0$, $i=1,2$, are the coefficients,
$r, s$ are nonnegative integers called thresholds with $r\leq s$,
and $\{R_t\}$ is a $\{0,1\}$-valued stochastic sequence with $R_t\in \mathcal{F}_{t-1}$.
Clearly, from (\ref{bpart}), it follows the recurrence relation
\begin{flalign*}
R_{t}=I(y_{t-1}\leq r)+I(r<y_{t-1}\leq s)R_{t-1},\quad t\in\mathbb{Z},
\end{flalign*}
where $I(\cdot)$ is the indicator function.


(b) A count time series $\{y_t: t\in\mathbb{Z}\}$ is said to follow a first-order HPART model,
if the intensity process $\lambda_t$ satisfies the stochastic recurrence 
\footnote{Note that there is more than one way to define the same condition. For example, the condition  `$\{\Delta y_{t-1} \geq c, y_{t-1} \leq s\}\hspace{2mm} \mathrm{or} \hspace{2mm}\{\Delta y_{t-1} < c, y_{t-1} \leq r\}$' \hspace{2mm} is equivalent to  the condition \hspace{2mm} `$\{y_{t-1} \leq r\}\hspace{2mm} \mathrm{or} \hspace{2mm} \{\Delta y_{t-1} > c, r < y_{t-1} \leq s\}$'.} 
\begin{flalign}\label{hpart}
\lambda_t =\left\{\begin{array}{lll}
\omega_1 + \alpha_1y_{t-1}+\beta_1\lambda_{t-1}, & \mathrm{if} \hspace{2mm} \{\Delta y_{t-1} \geq c, y_{t-1} \leq s\}\hspace{2mm} \mathrm{or} \hspace{2mm}\{\Delta y_{t-1} < c, y_{t-1} \leq r\},\\
\omega_2  + \alpha_2y_{t-1}+\beta_2\lambda_{t-1}, & \mathrm{if} \hspace{2mm} \{\Delta y_{t-1} \geq c, y_{t-1} > s\} \hspace{2mm} \mathrm{or} \hspace{2mm}\{\Delta y_{t-1} < c, y_{t-1} > r\},
\end{array}
\right.
\end{flalign}
where 
$\Delta y_{t-1} =  y_{t-1}-y_{t-2}$, $\omega_i>0,$ $\alpha_i\geq0,$ $\beta_i\geq0$, $i=1,2$, are the coefficients,
$r$, $s$, $c$ are integers called thresholds with $0\leq r\leq s$ and $c\in\mathbb{Z}$. For the HPART model, we have the following recurrence-free relations
\begin{flalign*}
I_t:=& I(\Delta y_{t-1} \geq c, y_{t-1} \leq s) + I(\Delta y_{t-1} < c, y_{t-1} \leq r)\\
=& I(y_{t-1} \leq r) + I(r < y_{t-1} \leq s) I(\Delta y_{t-1} < c).
\end{flalign*}

It is interesting to compare the role of $I(\Delta y_{t-1} < c)$ in the HPART model to the role of  $R_{t-1}$ in the BPART model. (i) The HPART model switches data in the hysteretic regime $\{r < y_{t-1} \leq s\}$ at different thresholds: at $r$ if $\Delta y_{t-1} < c$, but at
$s$ if $\Delta y_{t-1} \geq c$. (ii) The BPART model does not switch regime for data in the buffer regime $\{r < y_{t-1} \leq s\}$. (iii) While $\Delta y_{t-1} $ for the HPART model   only depends on $\{ y_{t-1}, y_{t-2} \}$, i.e. a 2-finite past, $R_t$ for the BPART model   depends on  $\{R_{t-i}, i>1\}$, i.e.
$\{y_{t-i}, i>1 \}$, an infinite past. 



\subsection{Main contributions}

There are four main contributions.

 (1) Motivated by the genuinely and physically meaningful notion of hysteresis, we propose a new nonlinear mechanism, the HPART model, for the intensity process of the parametric dynamic stochastic Poisson models, besides re-analysing the buffer mechanism of the BPART model. 

(2) For these two models, we discuss, in a unified manner, their maximum likelihood estimators (MLE)
 and establish their strong consistency and asymptotic normality.

(3) As the two models are separate in the sense of \cite{Cox1960,Cox1962}, we develop two separate tests for a BPART/HPART model vs. a HPART/BPART model and derive their respective limiting distributions. Specifically, we indicate that the test statistic under the BPART model can be represented as a function of multiple chi-square distributions, while that under the HPART model approximately follows a chi-square distribution. We validate our results by Monte Carlo simulation. 


(4) We use the escape custody data and the hepatitis B data to illustrate model selections, model fitting, and prediction performance of the BPART model and the HPART model.

\subsection{Outline} 
The remainder of the paper is organized as follows. For the BPART model and the HPART model, Section \ref{mle} considers, in a unified manner, the maximum likelihood estimation of the first-order case and establishes their asymptotic properties under general conditions.
Section \ref{testing} studies separate families of hypotheses and develops two tests for a BPART/HPART model vs. a HPART/BPART model.
Section \ref{simulation} reports results of 
simulation studies.
Section \ref{application} showcases two real applications. Section \ref{conclusion} concludes.
Proofs of all theorems with some technical lemmas are relegated to the Supplementary Material.

\section{Maximum Likelihood Estimation}\label{mle}
Let $\vartheta=(\omega_1, \alpha_1, \beta_1, \omega_2, \alpha_2, \beta_2)^T$ and
$\theta^{\mathrm{b}}=(\vartheta^T, r, s)^T$
and $\theta^{\mathrm{h}}=(\vartheta^T, r, s, c)^T$, $\tau_{\mathrm{b}}=(r, s)^T\in\mathbb{Z}^2$ and $\tau_{\mathrm{h}}=(r, s, c)^T\in\mathbb{Z}^3$.
Supposing that the observations $\{y_1,\ldots,y_{n}\}$ are available
with the true parameter $\theta_0$.
The conditional log-likelihood function is given by
\begin{flalign*}
\widetilde{L}_n(\theta)=\sum_{t=1}^n\tilde{\ell}_t(\theta) \quad\mbox{with} \quad
\tilde{\ell}_t(\theta)=-\widetilde{\lambda}_t(\theta)+y_t\log\widetilde{\lambda}_t(\theta)-\log(y_t!).
\end{flalign*}
The maximum likelihood estimator (MLE) of $\theta_{0}$ is defined as
\begin{flalign*}
\widehat{\theta}_{n}=\mathop{\arg\max}_{\theta\in \Theta_{\vartheta}\times\Theta_{\tau}}\widetilde{L}_n(\theta),
\end{flalign*}
where $\Theta_{\vartheta}$ is the coefficient parameter space, and $\Theta_{\tau}$ is
the threshold parameter space.

For the BPART model,
\begin{flalign*}
\widetilde{\lambda}_t(\theta)&=\{\omega_1+\alpha_1 y_{t-1}+\beta_1\widetilde{\lambda}_{t-1}(\theta)\}\widetilde{R}_t(\theta)
+\{\omega_2+\alpha_2 y_{t-1}+\beta_2\widetilde{\lambda}_{t-1}(\theta)\}[1-\widetilde{R}_t(\theta)],\\
\widetilde{R}_t(\theta)&=I(y_{t-1}\leq r)+I(r<y_{t-1}\leq s)\widetilde{R}_{t-1}(\theta),\qquad t\geq 1,
\end{flalign*}
where the initial value $(y_0, \widetilde{\lambda}_{0}(\theta), \widetilde{R}_0(\theta))^T$ is fixed.

For the HPART model,
\begin{flalign*}
\widetilde{\lambda}_t(\theta)&=\{\omega_1+\alpha_1 y_{t-1}+\beta_1\widetilde{\lambda}_{t-1}(\theta)\}I_t(\theta)
+\{\omega_2+\alpha_2 y_{t-1}+\beta_2\widetilde{\lambda}_{t-1}(\theta)\}[1-I_t(\theta)],\\
I_t(\theta)&=I(\{\Delta y_{t-1} \geq c, y_{t-1} \leq s\} \cup \{\Delta y_{t-1} < c, y_{t-1} \leq r\})\\
&=I(\Delta y_{t-1} \geq c)I(y_{t-1} \leq s)+I(\Delta y_{t-1} < c)I(y_{t-1} \leq r), \qquad t\geq 1, 
\end{flalign*}
where the initial value $(y_0, \widetilde{\lambda}_{0}(\theta), \Delta y_{0})^T$ is fixed.

To facilitate the study on the asymptotic properties of $\widehat{\theta}_n$, we define the theoretical log-likelihood function as
\begin{flalign*}
L_n(\theta)=\sum_{t=1}^n\ell_t(\theta) \quad\mbox{with} \quad
\ell_t(\theta)=-\lambda_t(\theta)+y_t\log\lambda_t(\theta)-\log(y_t!),
\end{flalign*}

For the BPART model,
\begin{flalign*}
\lambda_t(\theta)&=\{\omega_1+\alpha_1 y_{t-1}+\beta_1\lambda_{t-1}(\theta)\}R_t(\theta)
+\{\omega_2+\alpha_2 y_{t-1}+\beta_2\lambda_{t-1}(\theta)\}[1-R_t(\theta)],\\
R_t(\theta)&=I(y_{t-1}\leq r)+I(r<y_{t-1}\leq s)R_{t-1}(\theta),\qquad t\in\mathbb{Z}.
\end{flalign*}

For the HPART model,
\begin{flalign*}
\lambda_t(\theta)&=\{\omega_1+\alpha_1 y_{t-1}+\beta_1\lambda_{t-1}(\theta)\}I_t(\theta)
+\{\omega_2+\alpha_2 y_{t-1}+\beta_2\lambda_{t-1}(\theta)\}[1-I_t(\theta)],\\
I_t(\theta)&=I(\Delta y_{t-1} \geq c)I(y_{t-1} \leq s)+I(\Delta y_{t-1} < c)I(y_{t-1} \leq r), \qquad t\in\mathbb{Z}.
\end{flalign*}

To study the asymptotic properties of $\widehat{\theta}_n$, we need the following assumptions, 
which are standard in the study of integer-valued time series.

\begin{assumption}\label{paraspace}
The coefficient parameter space $\Theta_{\vartheta}\subseteq\{\vartheta\in\mathbb{R}^6|
\omega_i>0, \alpha_i>0, 0<\beta_i<1, \alpha_2+\beta_2<1, i=1, 2\}$ is compact. Let $\Theta_{\tau}$ denote the threshold parameter space, specifically 
$\Theta_{\tau}:=\Theta_{\tau_\mathrm{b}}=\{(r, s)^T\in \mathbb{Z}^2|~0\leq r<s\}$ for the BPART model, and 
$\Theta_{\tau}:=\Theta_{\tau_\mathrm{h}}=\{(r, s, c)^T\in \mathbb{Z}^3|~0\leq r<s\}$ for the HPART model. Assume that $\Theta_{\tau}$ is bounded.
\end{assumption}

\begin{assumption}\label{identification}
The true coefficient parameter satisfies $(\omega_{10}, \alpha_{10}, \beta_{10})^T\neq(\omega_{20}, \alpha_{20}, \beta_{20})^T$.
\end{assumption}

\begin{assumption}\label{s-e}
The sequence $\{y_t\}$ is strictly stationary and ergodic.
\end{assumption}

\begin{assumption}\label{interior}
The true coefficient parameter $\vartheta_0$ is an interior point of $\Theta_\vartheta$.
\end{assumption}

\begin{theorem}\label{consistency}
Under Assumptions \ref{paraspace}-\ref{s-e}, it follows that
$\widehat{\theta}_n\rightarrow\theta_0$ a.s. as $n\rightarrow\infty$.
\end{theorem}

Since thresholds $r_0$, $s_0$ and $c_0$ are integer-valued, the consistency of $\widehat{\tau}_n$ implies that $\widehat{r}_n=r_0$, $\widehat{s}_n=s_0$ and $\widehat{c}_n=c_0$ eventually. Thus, without loss of generality, we can assume that the parameter $\tau$ is known when we study the limiting distribution of 
the MLE $\widehat{\vartheta}_n$.

\begin{theorem}\label{normality}
Under Assumptions \ref{paraspace}-\ref{interior}, then
\begin{flalign*}
\sqrt n\,(\widehat{\vartheta}_n-\vartheta_0) \xrightarrow{\ d\ } N(0,G^{-1}),
\end{flalign*}
where $\xrightarrow{\ d\ }$ stands for convergence in distribution and
\begin{flalign}\label{SME}
G = E\Big\{\frac{1}{\lambda_t(\theta_0)}\frac{\partial\lambda_t(\theta_0)}{\partial\vartheta}
\frac{\partial\lambda_t(\theta_0)}{\partial\vartheta^T}\Big\}.
\end{flalign}
\end{theorem}

In order to construct a confidence region of the parameter $\vartheta_0$, it is necessary to estimate
the matrix $G$. A weakly consistent estimate is
\begin{flalign*}
\widehat{G}=n^{-1}\sum_{t=1}^n \frac{1}{\widetilde{\lambda}_t(\theta)}\frac{\partial\widetilde{\lambda}_t(\theta)}{\partial\vartheta}
\frac{\partial\widetilde{\lambda}_t(\theta)}{\partial\vartheta^T}\Bigg|_{\theta=\widehat{\theta}_n}.
\end{flalign*}

\section{Tests}\label{testing}
From models (\ref{bpart}) and (\ref{hpart}), we consider a coupled Poisson model with a compound intensity process,
namely
\begin{flalign*}
\mathcal{L}(y_t\mid\mathcal{F}_{t-1})=\mathrm{Poisson}(\lambda_t),\quad t\in\mathbb{Z},
\end{flalign*}
where $\lambda_t$ is a function of a weight factor $\delta$ and the parameter $\theta = (\vartheta^T, r, s, c)^T$, defined as
\begin{flalign}\label{compound}
\lambda_t=(1-\delta)\lambda_t^{\mathrm{b}}+\delta\lambda_t^{\mathrm{h}},\quad t\in\mathbb{Z}.
\end{flalign}
To ensure positivity of $\lambda_t$, $\delta\in[0, 1]$ is required.

\subsection{Test for a BPART model against a HPART model}
The following test 
\begin{flalign}\label{test:bvh}
H_0: \delta=0~~\mbox{against}~~H_a: \delta\neq0
\end{flalign}
tests the departure of a BPART model in the direction of model (\ref{compound}).

The conditional log-likelihood function is given by
\begin{flalign*}
&\widetilde{L}_n(\delta, \theta)=\sum_{t=1}^n\tilde{\ell}_t(\delta, \theta) \quad\mbox{with} \quad
\tilde{\ell}_t(\delta, \theta)=-\widetilde{\lambda}_t(\delta, \theta)+y_t\log\widetilde{\lambda}_t(\delta, \theta)-\log(y_t!),\\
&\hspace{30mm}\widetilde{\lambda}_t(\delta, \theta)=(1-\delta)\widetilde{\lambda}_t^{\mathrm{b}}(\theta)
+\delta\widetilde{\lambda}_t^{\mathrm{h}}(\theta).
\end{flalign*}
Then, we get the first and second derivatives for $\delta$ as below
\begin{flalign*}
\frac{\partial\widetilde{L}_n(\delta, \theta)}{\partial\delta}&=\sum_{t=1}^n\frac{\partial\tilde{\ell}_t(\delta, \theta)}{\partial\delta} =\sum_{t=1}^n\Big(\frac{y_t}{\widetilde{\lambda}_t(\delta, \theta)}-1\Big)\frac{\partial\widetilde{\lambda}_t(\delta, \theta)}{\partial\delta} =\sum_{t=1}^n\Big(\frac{y_t}{\widetilde{\lambda}_t(\delta, \theta)}-1\Big)\{\widetilde{\lambda}_t^{\mathrm{h}}(\theta)-\widetilde{\lambda}_t^{\mathrm{b}}(\theta)\},\\
\frac{\partial^2\widetilde{L}_n(\delta, \theta)}{\partial\delta^2}&=-\sum_{t=1}^n\frac{y_t}{\widetilde{\lambda}_t^2(\delta, \theta)}\{\widetilde{\lambda}_t^{\mathrm{h}}(\theta)-\widetilde{\lambda}_t^{\mathrm{b}}(\theta)\}^2.
\end{flalign*}
In model (\ref{bpart}), the threshold $c$ is a nuisance parameter. Under $H_0$, we fix $c \in \mathbb{Z}$ and the score function and the information matrix are 
\begin{flalign*}
\frac{\partial\widetilde{L}_n(0, \theta)}{\partial\delta}
=\sum_{t=1}^n\Big(\frac{y_t}{\widetilde{\lambda}_t^{\mathrm{b}}(\theta)}-1\Big)\{\widetilde{\lambda}_t^{\mathrm{h}}(\theta)-\widetilde{\lambda}_t^{\mathrm{b}}(\theta)\},
\end{flalign*}
\begin{flalign*}
\frac{\partial^2\widetilde{L}_n(0, \theta)}{\partial\delta^2}=-\sum_{t=1}^n\frac{y_t}{\{\widetilde{\lambda}_t^{\mathrm{b}}(\theta)\}^2}
\big\{\widetilde{\lambda}_t^{\mathrm{h}}(\theta)-\widetilde{\lambda}_t^{\mathrm{b}}(\theta)\big\}^2.
\end{flalign*}
For a given $c$, let
\begin{flalign*}
T^{\mathrm{b}}_n(c)=\Big\{-\frac{\partial^2\widetilde{L}_n(0, \theta^{\mathrm{b}},c)}{\partial\delta^2}\Big\}^{-1}
\Big\{\frac{\partial\widetilde{L}_n(0, \theta^{\mathrm{b}},c)}{\partial\delta}\Big\}^2\Big|_{\theta^{\mathrm{b}}=\widehat{\theta}_n^{\,\mathrm{b}}}.
\end{flalign*}
Assume the range of $c$ is $\{c_1, c_2,...,c_k\}$. Denote $[k] = \{1,2,...,k\}$ and 
\begin{flalign*}
\sigma_1(c_i,c_j)& = E\Big[\frac{1}{\lambda_t^{\mathrm{b}}(\theta_0)}\{\lambda_t^{\mathrm{h}}(\vartheta_0, r_0, s_0, c_i)-\lambda_t^{\mathrm{b}}(\theta_0)\}\{\lambda_t^{\mathrm{h}}(\vartheta_0, r_0, s_0, c_j)-\lambda_t^{\mathrm{b}}(\theta_0)\} \Big],\\
\sigma_2(c_i,c_j) &= \sigma_1(c_i,c_j)- E\Big[ \frac{1}{\lambda_t^{\mathrm{b}}(\theta_0)}\{\lambda_t^{\mathrm{h}}(\vartheta_0, r_0, s_0, c_i)-\lambda_t^{\mathrm{b}}(\theta_0)\}\frac{\partial \lambda_t^{\mathrm{b}}\left(\theta_0\right)}{\partial \vartheta^T}\Big]\\
&\quad \times\Sigma_1^{-1}E\Big[ \frac{1}{\lambda_t^{\mathrm{b}}(\theta_0)}\{\lambda_t^{\mathrm{h}}(\vartheta_0, r_0, s_0, c_j)-\lambda_t^{\mathrm{b}}(\theta_0)\}\frac{\partial \lambda_t^{\mathrm{b}}\left(\theta_0\right)}{\partial \vartheta}\Big], \hspace{2mm} i,j \in [k],
\end{flalign*}
where $\Sigma_1 = G\big|_{\lambda_t=\lambda_t^{\mathrm{b}}}$ and $G$ is defined in (\ref{SME}).

The score-based test statistic for testing BPART against HPART is defined as
\begin{flalign*}
S_{n}= \max_{i \in [k]}T_{n}^{\mathrm{b}}(c_i).
\end{flalign*}
\begin{theorem}\label{thm:bvh}
Under $H_0$, if Assumptions \ref{paraspace}-\ref{interior} hold, then
\begin{flalign*}
    S_{n} \xrightarrow{\ d\ } \max_{i \in [k]} \frac{Z^2(c_i)}{\sigma_1(c_i,c_i)},
\end{flalign*}
where the random vector $(Z(c_1),...,Z(c_k))^T \sim N(0,\Sigma)$ with $\Sigma = (\sigma_2(c_i,c_j))_{k\times k}$.
\end{theorem}

\subsection{Test for a HPART model against a BPART model}
The following test
\begin{flalign}\label{test:hvb}
\widetilde{H}_0: \delta=1~~\mbox{against}~~\widetilde{H}_a: \delta\neq1
\end{flalign}
tests the departure of a HPART model in the direction of model (\ref{compound}).
Under $\widetilde{H}_0$, we obtain the score function and information matrix as
follows
\begin{flalign*}
\frac{\partial\widetilde{L}_n(1, \theta)}{\partial\delta}
&=\sum_{t=1}^n\Big(\frac{y_t}{\widetilde{\lambda}_t^{\mathrm{h}}(\theta)}-1\Big)\{\widetilde{\lambda}_t^{\mathrm{h}}(\theta)-\widetilde{\lambda}_t^{\mathrm{b}}(\theta)\},\\
\frac{\partial^2\widetilde{L}_n(1, \theta)}{\partial\delta^2}&=-\sum_{t=1}^n\frac{y_t}{\{\widetilde{\lambda}_t^{\mathrm{h}}(\theta)\}^2}
\big\{\widetilde{\lambda}_t^{\mathrm{h}}(\theta)-\widetilde{\lambda}_t^{\mathrm{b}}(\theta)\big\}^2.
\end{flalign*}
The score-based test statistic under $\widetilde{H}_0$ is
\begin{flalign*}
T^{\mathrm{h}}_n=\Big\{-\frac{\partial^2\widetilde{L}_n(1, \theta)}{\partial\delta^2}\Big\}^{-1}
\Big\{\frac{\partial\widetilde{L}_n(1, \theta)}{\partial\delta}\Big\}^2\Big|_{\theta=\widehat{\theta}_n^{\,\mathrm{h}}}
\end{flalign*}
Similarly, denote
\begin{flalign*}
&\sigma_1' = E\Big\{ -\frac{1}{n}\frac{\partial^2 L\left(1, {\theta}_0\right)}{\partial \delta^2}\Big\} = E\Big[\frac{1}{\lambda_t^{\mathrm{h}}(\theta_0)}\{\lambda_t^{\mathrm{h}}(\theta_0)-\lambda_t^{\mathrm{b}}(\theta_0)\}^2 \Big],\\
&\sigma_2' = \sigma_1'- E\Big[ \frac{1}{\lambda_t^{\mathrm{h}}(\theta_0)}\{\lambda_t^{\mathrm{h}}(\theta_0)-\lambda_t^{\mathrm{b}}(\theta_0)\}\frac{\partial \lambda_t^{\mathrm{h}}\left(\theta_0\right)}{\partial \vartheta^T}\Big]\Sigma_2^{-1}E\Big[ \frac{1}{\lambda_t^{\mathrm{h}}(\theta_0)}\{\lambda_t^{\mathrm{h}}(\theta_0)-\lambda_t^{\mathrm{b}}(\theta_0)\}\frac{\partial \lambda_t^{\mathrm{h}}\left(\theta_0\right)}{\partial \vartheta}\Big],
\end{flalign*}
where $\Sigma_2 = G\big|_{\lambda_t=\lambda_t^{\mathrm{h}}}$ and $G$ is defined in (\ref{SME}).
\begin{theorem}\label{thm:hvb}
Under $\widetilde{H}_0$, if Assumptions \ref{paraspace}-\ref{interior} hold, then 
\begin{flalign*}
T_{n}^{\mathrm{h}} \frac{\hat{\sigma}_1'}{\hat{\sigma}_2'}\xrightarrow{\ d\ } \chi_1^2.
\end{flalign*}
\end{theorem}


\section{Simulation Studies}\label{simulation}
\subsection{Performance of estimators}
To evaluate the finite-sample performance of $\widehat{\theta}_n$, we take two sets of parameters for Monte Carlo simulation of the two models.

For the BPART model, observations are generated from the model with true parameters
$\theta^{\mathrm{b}}_0=(0.5, 0.6, 0.4, 0.2, 0.4, 0.5, 3, 6)^T$ and $\theta^{\mathrm{b}}_0=(0.6, 0.8, 0.7, 0.4, 0.2, 0.2, 4, 7)^T$, respectively. 

For the HPART model, observations are generated from the model with true parameters
$\theta^{\mathrm{h}}_0=(0.5, 0.6, 0.4, 0.2, 0.4, 0.5, 3, 6,0)^T$ and $\theta^{\mathrm{h}}_0=(0.6, 0.8, 0.7, 0.4, 0.2, 0.2, 4, 7,-1)^T$, respectively.
We adopt the same parameters $\vartheta_0$ in the two models in order to better learn about the performance in parameter estimation.

The length of observations 
is $n=500$, 1000, and 2000, respectively.
The tables below summarize the empirical mean (EM),
the empirical variance $\mathrm{cov}(\widehat{\vartheta}_n)$ (EV), the sample mean of $\widehat{G}^{-1}/n$ (SG) and the ratios $EV/EM$ (V/M) based on 1000 replications.

\subsubsection{BPART model}
\begin{table}[!htbp]
  \renewcommand{\arraystretch}{0.7} 
  \centering
  \caption{Summary of simulation results for the first set of true parameters(BPART)}
  \begin{tabular}{lccccccccc}
    \toprule
    $n$ &\multicolumn{1}{c}{Description}&
        \multicolumn{1}{c}{$\omega_1$}&
        \multicolumn{1}{c}{$\alpha_1$}&
        \multicolumn{1}{c}{$\beta_1$}&
        \multicolumn{1}{c}{$\omega_2$}&
        \multicolumn{1}{c}{$\alpha_2$}&
        \multicolumn{1}{c}{$\beta_2$}&
        \multicolumn{1}{c}{$r$}&
        \multicolumn{1}{c}{$s$}\\
    \midrule
    &$\theta_0^{\mathrm{b}}$ &0.5 &0.6 &0.4 &0.2 &0.4 &0.5 &3 &6 \\
    \hline
    $500$   &EM &0.586 &0.600 &0.382 &0.373 &0.392 &0.476 &3.051 &6.042 \\
    &EV &0.095 &0.007 &0.009 &0.176 &0.006 &0.012  &0.413 &0.355\\
    &SG &0.082 &0.005 &0.008 &0.395 &0.006 &0.014 &/ &/ \\
    &V/M &0.163 &0.012 &0.022 &0.471 &0.016 &0.025 &0.135 &0.059\\
    \hline
    $1000$   &EM &0.543 &0.598 &0.392 &0.289 &0.395 &0.490 &3.022 &6.004 \\
    &EV &0.040 &0.003 &0.004 &0.069 &0.003 &0.005 &0.108 &0.060 \\
    &SG &0.035 &0.002 &0.004 &0.153 &0.003 &0.006 &/ &/ \\
    &V/M &0.074 &0.005 &0.010 &0.238 &0.007 &0.010 &0.036 &0.010\\
    \hline
        $2000$   &EM &0.516 &0.598 &0.399 &0.246 &0.398 &0.495 &2.998 &6.001 \\
    &EV &0.016 &0.001 &0.002 &0.027 &0.001 &0.002 &0.008 &0.001 \\
    &SG &0.016 &0.001 &0.002 &0.069 &0.001 &0.003 &/ &/ \\
    &V/M &0.031 &0.002 &0.004 &0.111 &0.003 &0.004 &0.003 &0.000\\
    \bottomrule
  \end{tabular}\label{1BPART:s1}
\end{table}
\begin{table}[!htbp]
  \renewcommand{\arraystretch}{0.7}
  \centering
  \caption{Summary of simulation results for the second set of true parameters(BPART)}
  \begin{tabular}{lccccccccc}
    \toprule
    $n$ &\multicolumn{1}{c}{Description}&
        \multicolumn{1}{c}{$\omega_1$}&
        \multicolumn{1}{c}{$\alpha_1$}&
        \multicolumn{1}{c}{$\beta_1$}&
        \multicolumn{1}{c}{$\omega_2$}&
        \multicolumn{1}{c}{$\alpha_2$}&
        \multicolumn{1}{c}{$\beta_2$}&
        \multicolumn{1}{c}{$r$}&
        \multicolumn{1}{c}{$s$}\\
    \midrule
    &$\theta_0^{\mathrm{b}}$ &0.6 &0.8 &0.7 &0.4 &0.2 &0.2 &4 &7 \\
    \hline
    $500$   &EM &0.609 &0.792 &0.707 &0.462 &0.192 &0.203 &4 &7 \\
    &EV &0.078 &0.008 &0.008 &0.099 &0.003 &0.004 &0 &0 \\
    &SG &0.093 &0.007 &0.008 &0.166 &0.004 &0.004 &/ &/ \\
    &V/M &0.128 &0.010 &0.011 &0.215 &0.018 &0.022 &0 &0\\
    \hline
    $1000$   &EM &0.609 &0.797 &0.702 &0.438 &0.195 &0.201 &4 &7 \\
    &EV &0.042 &0.004 &0.004 &0.057 &0.002 &0.002 &0 &0 \\
    &SG &0.046 &0.004 &0.004 &0.083 &0.002 &0.002 &/ &/ \\
    &V/M &0.069 &0.005 &0.006 &0.130 &0.009 &0.011 &0 &0\\
    \hline
    $2000$   &EM &0.613 &0.797 &0.700 &0.420 &0.198 &0.200 &4 &7 \\
    &EV &0.022 &0.002 &0.002 &0.033 &0.001 &0.001 &0 &0 \\
    &SG &0.023 &0.002 &0.002 &0.041 &0.001 &0.001 &/ &/ \\
    &V/M &0.035 &0.003 &0.003 &0.079 &0.005 &0.005 &0 &0\\
    \bottomrule
  \end{tabular}
  \label{1BPART:s2}
\end{table}

\FloatBarrier
\subsubsection{HPART model}
\begin{table}[!htbp]
  \renewcommand{\arraystretch}{0.7}
  \centering
  \caption{Summary of simulation results for the first set of true parameters(HPART)}
  \begin{tabular}{lcccccccccc}
    \toprule
   $n$ &\multicolumn{1}{c}{Description}&
        \multicolumn{1}{c}{$\omega_1$}&
        \multicolumn{1}{c}{$\alpha_1$}&
        \multicolumn{1}{c}{$\beta_1$}&
        \multicolumn{1}{c}{$\omega_2$}&
        \multicolumn{1}{c}{$\alpha_2$}&
        \multicolumn{1}{c}{$\beta_2$}&
        \multicolumn{1}{c}{$r$}&
        \multicolumn{1}{c}{$s$}&
        \multicolumn{1}{c}{$c$}\\
    \midrule
    &$\theta_0^{\mathrm{h}}$ &0.5 &0.6 &0.4 &0.2 &0.4 &0.5 &3 &6 &0\\
    \hline
    $500$   &EM &0.620 &0.598 &0.376 &0.418 &0.394 &0.464 &3.144 &6.087 &0.068\\
    &EV &0.138 &0.012 &0.011 &0.375 &0.010 &0.018  &1.140 &0.704 &1.799\\
    &SG &0.093 &0.006 &0.008 &0.602 &0.006 &0.016 &/ &/ &/\\
    &V/M &0.222 &0.019 &0.030 &0.896 &0.025 &0.038 &0.363 &0.116 &26.46\\
    \hline
        $1000$   &EM &0.559 &0.597 &0.390 &0.313 &0.399 &0.479 &2.988 &5.994 &-0.106 \\
    &EV &0.053 &0.003 &0.005 &0.124 &0.003 &0.006 &0.681 &0.114 &0.691\\
    &SG &0.040 &0.003 &0.004 &0.230 &0.003 &0.007 &/ &/ &/\\
    &V/M &0.094 &0.005 &0.012 &0.396 &0.009 &0.013 &0.228 &0.019 &-6.523\\
    \hline
        $2000$   &EM &0.525 &0.597 &0.398 &0.260 &0.398 &0.491 &2.898 &6 &-0.063\\
    &EV &0.022 &0.001 &0.002 &0.037 &0.001 &0.002 &0.292 &0 &0.157 \\
    &SG &0.018 &0.001 &0.002 &0.096 &0.001 &0.003 &/ &/ &/\\
    &V/M &0.042 &0.002 &0.005 &0.141 &0.003 &0.005 &0.101 &0 &-2.495\\
    \bottomrule
  \end{tabular}\label{1HAR:s1}
\end{table}

\begin{table}[!htbp]
  \renewcommand{\arraystretch}{0.7}
  \centering
  \caption{Summary of simulation results for the second set of true parameters(HPART)}
  \begin{tabular}{lcccccccccc}
    \toprule
    $n$ &\multicolumn{1}{c}{Description}&
        \multicolumn{1}{c}{$\omega_1$}&
        \multicolumn{1}{c}{$\alpha_1$}&
        \multicolumn{1}{c}{$\beta_1$}&
        \multicolumn{1}{c}{$\omega_2$}&
        \multicolumn{1}{c}{$\alpha_2$}&
        \multicolumn{1}{c}{$\beta_2$}&
        \multicolumn{1}{c}{$r$}&
        \multicolumn{1}{c}{$s$}&
        \multicolumn{1}{c}{$c$}\\
    \midrule
    &$\theta_0^{\mathrm{h}}$ &0.6 &0.8 &0.7 &0.4 &0.2 &0.2 &4 &7 &-1\\
    \hline
    $500$   &EM &0.616 &0.792 &0.705 &0.433 &0.193 &0.204 &4 &7 &-1\\
    &EV &0.086 &0.007 &0.008 &0.100 &0.003 &0.005 &0 &0 &0\\
    &SG &0.100 &0.007 &0.008 &0.185 &0.004 &0.004 &/ &/ &/\\
    &V/M &0.139 &0.009 &0.011 &0.230 &0.017 &0.022 &0 &0 &0\\
    \hline
    $1000$   &EM &0.610 &0.797 &0.702 &0.426 &0.194 &0.204 &4 &7 &-1\\
    &EV &0.046 &0.004 &0.004 &0.063 &0.002 &0.002 &0 &0 &0\\
    &SG &0.049 &0.003 &0.004 &0.091 &0.002 &0.002 &/ &/ &/\\
    &V/M &0.075 &0.005 &0.006 &0.147 &0.009 &0.011 &0 &0 &0\\
    \hline
    $2000$   &EM &0.611 &0.798 &0.700 &0.410 &0.199 &0.200 &4 &7 &-1\\
    &EV &0.024 &0.002 &0.002 &0.031 &0.001 &0.001 &0 &0 &0\\
    &SG &0.024 &0.002 &0.002 &0.046 &0.001 &0.001 &/ &/ &/\\
    &V/M &0.039 &0.002 &0.003 &0.076 &0.004 &0.005 &0 &0 &0\\
    \bottomrule
  \end{tabular}
  \label{1HAR:s2}
\end{table}

From the Tables above, we can see that $\widehat{\tau}_n$ converges to
$\tau_0$ generally.
It seems that the speed of convergence 
heavily depends on other parameters. Specifically, corresponding to different sets of true parameters,
Tables \ref {1BPART:s1} and \ref {1HAR:s1} show  that even when $n$ is as large as 2000,
$\widehat{r}_n$ fails to hit $r_0$ exactly, while Tables \ref {1BPART:s2} and \ref {1HAR:s2} show that 
$(\widehat{r}_n,\widehat{s}_n)$ hits $(r_0,s_0)$ exactly when $n=500$. 
This phenomenon is also observed in the SETPAR model; see \cite{CW}.
Overall, the asymptotic results of the estimates of the  parameters
are confirmed in both examples. Specifically, the means of the estimates of the 
parameters are close to the true values of the parameters
and the accuracy increases as the length of observations increases. However, for the intercept
parameter $\omega_2$ estimate,  there is apparently a large variance-to-mean ratio in both examples, compared with the other parameter estimates.  This phenomenon also exists in the PAR  model, the SETPAR model, and even in the classical \textsc{GARCH}(1,1) model. So far, no explanation has been given in the literature.

\subsection{Performance of tests}
We evaluate the performance of the above tests 
through Monte Carlo experiments. Observations are generated 
with the same true parameters as above 
except that we set $c_0 = 0, -1,1$ for the study of power of the tests.
Here,
$n$ = 500, 1000, 2000 and 4000, respectively. We set the significance levels at 0.1, 0.05, and 0.01. Tables below summarize the size and power based on 1000
replications.

\subsubsection{Performance of test for the BPART model against the HPART model}

\begin{table}[!htbp]
  \renewcommand{\arraystretch}{0.7}
  \centering
  \caption{Summary of simulation results for the first set of true parameters in testing $H_0$}
  \begin{tabular}{lccccccc}
    \toprule
    &\multicolumn{1}{c}{Data}
    &\multicolumn{1}{c}{$c_0$}
    &\multicolumn{1}{c}{$\alpha$}
    &\multicolumn{1}{c}{$n=500$} &\multicolumn{1}{c}{$n=1000$}
    &\multicolumn{1}{c}{$n=2000$}&\multicolumn{1}{c}{$n=4000$}\\
    \midrule
    size &BPART & &0.10 &0.155 &0.127 &0.108 &0.105\\
    & & &0.05 &0.082 &0.074 &0.058 &0.062\\
    & & &0.01 &0.029 &0.018 &0.018 &0.013\\
    \hline
    power   &HPART &0 &0.10 &0.301 &0.459 &0.706 &0.867\\
    & & &0.05 &0.190 &0.327 &0.588 &0.807\\
    & & &0.01 &0.063 &0.158 &0.390 &0.662\\
    \hline
    power   &HPART &-1 &0.10 &0.323 &0.534 &0.835 &0.982\\
    & & &0.05 &0.202 &0.387 &0.740 &0.967\\
    & & &0.01 &0.086 &0.182 &0.501 &0.894\\
    \hline
    power   &HPART &1 &0.10 &0.364 &0.555 &0.789 &0.882\\
    & & &0.05 &0.241 &0.455 &0.719 &0.848\\
    & & &0.01 &0.105 &0.300 &0.602 &0.773\\
    \bottomrule
  \end{tabular}
  \label{1BvH:s1}
\end{table}

\begin{table}[!htbp]
  \renewcommand{\arraystretch}{0.7}
  \centering
  \caption{Summary of simulation results for the second set of true parameters in testing $H_0$}
  \begin{tabular}{lccccccc}
    \toprule
    &\multicolumn{1}{c}{Data}
    &\multicolumn{1}{c}{$c_0$}
    &\multicolumn{1}{c}{$\alpha$}
    &\multicolumn{1}{c}{$n=500$} &\multicolumn{1}{c}{$n=1000$}
    &\multicolumn{1}{c}{$n=2000$}&\multicolumn{1}{c}{$n=4000$}\\
    \midrule
    size &BPART & &0.10 &0.141 &0.123 &0.127 &0.130\\
    & & &0.05 &0.088 &0.064 &0.073 &0.067\\
    & & &0.01 &0.040 &0.021 &0.022 &0.014\\
    \hline
    power   &HPART &0 &0.10 &0.998 &1 &1 &1\\
    & & &0.05 &0.996 &1 &1 &1\\
    & & &0.01 &0.988 &1 &1 &1\\
    \hline
    power   &HPART &-1 &0.10 &0.989 &1 &1 &1\\
    & & &0.05 &0.986 &1 &1 &1\\
    & & &0.01 &0.968 &1 &1 &1\\
    \hline
    power   &HPART &1 &0.10 &1 &1 &1 &1\\
    & & &0.05 &1 &1 &1 &1\\
    & & &0.01 &1 &1 &1 &1\\
    \bottomrule
  \end{tabular}
  \label{1BvH:s2}
\end{table}

When testing $H_0$, the tables below show that the size and the power of the test approach their desired values under different scenarios, as $n$ increases. For example, 
Table \ref{1BvH:s2} shows satisfactory results in terms of 
power. However, the situation with  
Table \ref{1BvH:s1} is different. Although the size 
appears to be fairly well matched, the power behaviour is much less satisfactory before the sample reaches 4000. 
Further, values of $c_0$ appear to exert some effect on the power. 

\FloatBarrier
\subsubsection{Performance of test for the HPART model against the BPART model}
The results in Table \ref{1HvB:s1} and Table  \ref{1HvB:s2} for testing $\widetilde{H}_0$ are generally better than those for testing $H_0$ discussed previously.
\begin{table}[!htbp]
  \renewcommand{\arraystretch}{0.7}
  \centering
  \caption{Summary of simulation results for the first set of true parameters in testing $\widetilde{H}_0$}
  \begin{tabular}{lccccccc}
    \toprule
    &\multicolumn{1}{c}{Data}
    &\multicolumn{1}{c}{$c_0$}
    &\multicolumn{1}{c}{$\alpha$}
    &\multicolumn{1}{c}{$n=500$} &\multicolumn{1}{c}{$n=1000$}
    &\multicolumn{1}{c}{$n=2000$}&\multicolumn{1}{c}{$n=4000$}\\
    \midrule
    size   &HPART &0 &0.10 &0.087 &0.088 &0.088 &0.104\\
    & & &0.05 &0.040 &0.040 &0.045 &0.058\\
    & & &0.01 &0.015 &0.009 &0.009 &0.010\\
    \hline
    size   &HPART &-1 &0.10 &0.082 &0.085 &0.104 &0.104\\
    & & &0.05 &0.040 &0.051 &0.060 &0.050\\
    & & &0.01 &0.003 &0.006 &0.010 &0.009\\
    \hline
    size   &HPART &1 &0.10 &0.088 &0.102 &0.103 &0.113\\
    & & &0.05 &0.053 &0.046 &0.043 &0.053\\
    & & &0.01 & 0.009 &0.008 &0.008 &0.010\\
    \hline
    power   &BPART & &0.10 &0.296 &0.609 &0.868 &0.968 \\
    & & &0.05 &0.236 &0.544 &0.837 &0.955\\
    & & &0.01 &0.121 &0.407 &0.770 &0.932\\
    \bottomrule
  \end{tabular}
  \label{1HvB:s1}
\end{table}

\begin{table}[!htbp]
  \renewcommand{\arraystretch}{0.7}
  \centering
  \caption{Summary of simulation results for the second set of true parameters in testing $\widetilde{H}_0$}
  \begin{tabular}{lccccccc}
    \toprule
    &\multicolumn{1}{c}{Data}
    &\multicolumn{1}{c}{$c_0$}
    &\multicolumn{1}{c}{$\alpha$}
    &\multicolumn{1}{c}{$n=500$} &\multicolumn{1}{c}{$n=1000$}
    &\multicolumn{1}{c}{$n=2000$}&\multicolumn{1}{c}{$n=4000$}\\
    \midrule
    size   &HPART &0 &0.10 &0.095 &0.091 &0.109 &0.117\\
    & & &0.05 &0.063 &0.056 &0.057 &0.058\\
    & & &0.01 &0.017 &0.010 &0.015 &0.012\\
    \hline
    size   &HPART &-1 &0.10 &0.85 &0.096 &0.096 &0.112\\
    & & &0.05 &0.043 &0.047 &0.045 &0.055\\
    & & &0.01 &0.010 &0.010 &0.008 &0.006\\
    \hline
    size   &HPART &1 &0.10 &0.100 &0.094 &0.107 &0.110\\
    & & &0.05 &0.043 &0.052 &0.052 &0.053\\
    & & &0.01 & 0.009 &0.009 &0.015 &0.014\\
    \hline
    power   &BPART & &0.10 &0.989 &1 &1 &1\\
    & & &0.05 &0.988 &1 &1 &1\\
    & & &0.01 &0.972 &1 &1 &1\\
    \bottomrule
  \end{tabular}
  \label{1HvB:s2}
\end{table}

\FloatBarrier
\section{ Applications to Real Data}\label{application}
\subsection{Escape custody Data}
\begin{figure}[!htbp]
  \centering
  \includegraphics[width=0.9\linewidth]{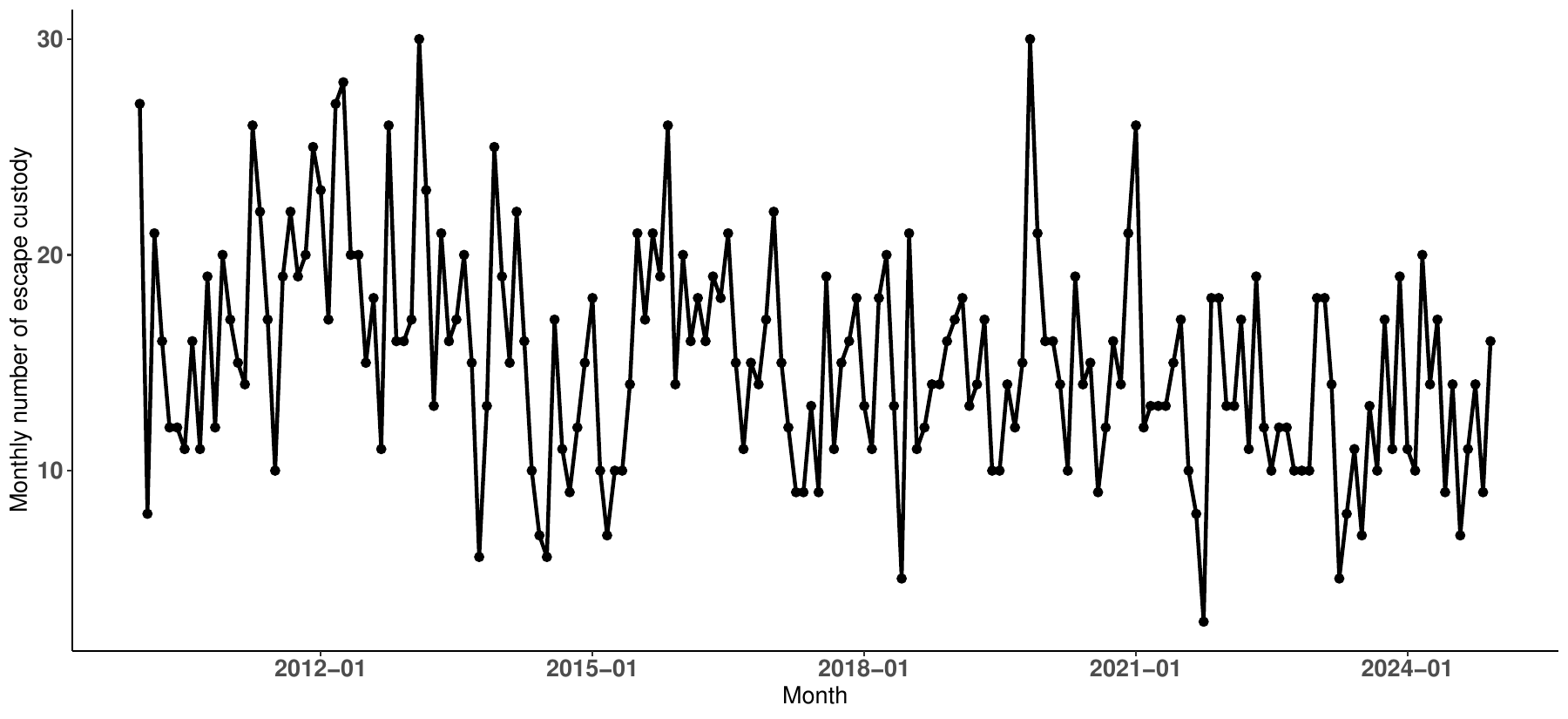}
  \caption{The monthly number of escape custody during
2010-2024.}
  \label{escape}
\end{figure}

To illustrate the efficacy of the two models and the tests, as well as to gain further insights, 
we analyse the monthly number of escape custody in the NSW during the period from 2010 to 2024. The data, available at https://bocsar.nsw.gov.au,  consists of 180 observations and is plotted in Fig.~\ref{escape}.
We use the first 160 data to fit models and the last 20 data to perform rolling out-of-sample one-step-ahead predictions.  
The fitted PAR model is 
\begin{flalign*}
\mathcal{L}(y_t\mid\mathcal{F}_{t-1})&=\mathrm{Poisson}(\lambda_t),\\
\lambda_t &= 2.42_{(1.42)} + 0.16_{(0.05)} y_{t-1}+0.68_{(0.12)}\lambda_{t-1},
\end{flalign*}
On noticing that the upward spikes tend to be steeper than the downward spikes, a SETPAR model might be worthy of consideration.  
The fitted SETPAR model is 
\begin{flalign*}
\mathcal{L}(y_t\mid\mathcal{F}_{t-1})&=\mathrm{Poisson}(\lambda_t),\\
\lambda_t &=
\begin{cases}
\begin{aligned}
&3.45 && + 0.04 y_{t-1} && + 0.70 \lambda_{t-1} \\[-1ex]
& (6.35) &&\quad (0.26) &&\quad (0.39)
\end{aligned} 
&\mathrm{if} \hspace{2mm} y_{t-1} \leq 11, \\[20pt]
\begin{aligned}
&1.61 && + 0.20 y_{t-1} && + 0.69 \lambda_{t-1} \\[-1ex]
& (1.66) &&\quad (0.08) &&\quad (0.14)
\end{aligned} 
&\mathrm{if} \hspace{2mm} y_{t-1} > 11.
\end{cases}
\end{flalign*}


Noticing that much more data lie within the middle range, we investigate the possibility of a BPART/HPART model. 
First, consider tests between the BPART model and the HPART model.
\begin{flalign}\label{BPART1}
H_0 : \quad\notag&\mathcal{L}(y_t\mid\mathcal{F}_{t-1})=\mathrm{Poisson}(\lambda_t),\\
&\lambda_t =
\begin{cases}
\begin{aligned}
&0.03 && + 0.24 y_{t-1} && + 0.82 \lambda_{t-1} \\[-1ex]
& (4.40) &&\quad (0.11) &&\quad (0.32)
\end{aligned} 
&\mathrm{if} \hspace{2mm} R_t = 1, \\[20pt]
\begin{aligned}
&0.007 && + 0.27 y_{t-1} && + 0.69 \lambda_{t-1} \\[-1ex]
& (1.66) &&\quad (0.08) &&\quad (0.12)
\end{aligned} 
&\mathrm{if} \hspace{2mm} R_t = 0.
\end{cases}\qquad
R_{t} =
	\begin{cases}
		1, & \mathrm{if} \hspace{2mm} y_{t-1}\leq 11,\\
		0, & \mathrm{if} \hspace{2mm}  y_{t-1}> 16,\\
		R_{t-1}, & \mathrm{otherwise},
	\end{cases}
\end{flalign}

\begin{flalign}\label{HPART1}
\widetilde{H}_0 : \quad\notag&\mathcal{L}(y_t\mid\mathcal{F}_{t-1})=\mathrm{Poisson}(\lambda_t),\\
&\lambda_t =
\begin{cases}
\begin{aligned}
&1.02 && + 0.24 y_{t-1} && + 0.75 \lambda_{t-1} \\[-1ex]
& (3.46) &&\quad (0.12) &&\quad (0.26)
\end{aligned} 
&\mathrm{if} \hspace{2mm} \{\Delta y_{t-1} \geq -2, y_{t-1} \leq 16\} \hspace{2mm} \mathrm{or} \hspace{2mm}   \{\Delta y_{t-1} < -2, y_{t-1} \leq 11\}, \\[20pt]
\begin{aligned}
&0.02 && + 0.39 y_{t-1} && + 0.53 \lambda_{t-1} \\[-1ex]
& (2.63) &&\quad (0.10) &&\quad (0.17)
\end{aligned} 
&\mathrm{if} \hspace{2mm} \{\Delta y_{t-1} \geq -2, y_{t-1} > 16\}\hspace{2mm} \mathrm{or} \hspace{2mm} \{\Delta y_{t-1} < -2, y_{t-1} > 11\}.
\end{cases}
\end{flalign}

Before doing so,  a preliminary examination of the four models reveals the following features: (i) the four models share a more or less similar coefficient for $\lambda_{t-1}$, implying that the difference of their $\lambda_t$ values is determined to a large extent by the combined effect of the intercept and $y_{t-1}$; (ii) on setting $y_{t-1}$ at its sample mean, 
15.17, 
the combined effect is roughly 4.8 for the PAR model; (iii) for the SETPAR model, similar 
calculations suggest that the combined effect (now regime dependent) is roughly 
4.0  for the lower regime with sample mean 13.64,  and 
4.8  for the upper regime with sample mean 15.71. Since there are much fewer data points in the lower regime, 
the  combined effect is roughly the same as for the PAR model; (iv) for the BPART model, the intercepts are negligible, implying that the key influencer is $y_{t-1}$ with a similar coefficient in the two regimes, leading to combined effects of roughly 
3.4 for the lower regime with sample mean 14.19, and 
4.3 for the upper regime with sample mean 15.82, but the number 
3.4 cannot be dismissed because the buffer regime recruits more data points into the lower regime than it is the case for the SETPAR model; (v) for the HPART model, similar 
calculations lead to combined effects of 
4.5 for the lower regime with the sample mean 14.35, and 
6.2 for the upper regime with the sample mean 15.93; (vi) overall, the combined effects on the PAR model and the SETPAR model are similar, but for the BPART model and the HPART model, the combined effects are clearly regime dependent, affecting the two models quite differently.      

Now, from Table \ref{test result1}, when testing $H_0$, we do not reject model (\ref{BPART1}) at each of the three levels. When testing $\widetilde{H}_0$, again we do not reject the model (\ref{HPART1}) at each of the three levels. These test results mean that we are not in a position to prefer one model to the other as far as model fitting is concerned. 
\begin{table}[!htbp]
  \centering
  \caption{Test results}
  \begin{tabular}{lrrr}
    \toprule
    \quad  &$\alpha = 0.1$ &$\alpha = 0.05$ &$\alpha = 0.01$ \\
    \midrule
    Test (\ref{BPART2}) &Not rejected &Not rejected &Not rejected \\
    Test (\ref{HPART2}) &Not rejected &Not rejected &Not rejected \\
    \bottomrule
  \end{tabular}
  \label{test result1}
\end{table}

In the BPART model, the intercepts ($\omega_1\approx0$, $\omega_2\approx0$) represent the baseline escape rate when past counts and past conditional means are negligible. The higher value of $\beta_1 = 0.82$ in the lower regime suggests that the process remains at a low level for a longer period, reflecting a more stable and persistent pattern. In contrast, a smaller $\beta_2$ than $\beta_1$ implies that once a large number is observed, the conditional mean of the process decreases more quickly compared to the lower regime. This suggests that the monthly number of escape custody cases is unlikely to  increase unduly
reflecting a relatively stable situation. The number of escape custody cases reached its minimum of 3 only in one month, namely October 2021, and the maximum value of 30 was observed only in two months, namely February 2013 and November 2021. In fact, although the maximum minus the minimum is as large as 27, the 80th percentile (at 19) minus the 20th percentile (at 11) is only 8. Therefore, the monthly counts exhibit a relatively stable and remain mostly within a mid-range 
over time. The HPART model leads generally to 
a similar conclusion, with minor variations: (i) $\omega_1 = 1.02$ indicates that a small number of escape custody cases in a given month tends to lead to an increased number of cases in the subsequent month. (ii) The relatively small magnitude of $c$ suggests that within the range $(11,16)$, the fluctuations in $y_t$ are mild.

For further comparison, we turn to out-of-sample prediction. The results are given in Table \ref{prediction1}, which includes additionally the PAR model and the SETPAR model for the sake of comparison. According to both the mean square error (MSE) and the mean absolute error (MAE), the HPART model outperforms the PAR model by about 3.7\%  and 2\% respectively and the BPART model less so. This result reveals that incorporating a hysteretic/buffer regime is potentially beneficial, to a more or less extent, within the context of nonlinearity and prediction performance.

\begin{table}[!htbp]
  \centering
  \caption{Summary of prediction error}
  \begin{tabular}{lrrrr}
    \toprule
    \quad  &PAR &SETPAR &BPART &HPART\\
    \midrule
    MSE &16.89 &17.39 &16.80 &16.27\\
    MAE &3.55 &3.60 &3.50 &3.47\\
    \bottomrule
  \end{tabular}
  \label{prediction1}
\end{table}

Since the fitted HPART model and the fitted BPART model are very close in terms of model parameters and prediction performance, it is relevant to probe deeper. 
Notice that both models allocate 
each datum to one of two regimes. Let us label them regime 
$1$ (the upper regime) and regime 
$0$ (the lower regime). So, each datum carries an identity (ID) card that bears either a 
$1$ marking or an 
$0$  marking. Of  the first 160 data, 60 of them are in the same buffer/hysteretic regime and the ID cards of 46 of them bear the same 
$1$ or $0$ marking .  That is to say a high proportion of the data in the  buffer regime of the BPART model is assigned to the same 
upper or lower regime as data in the hysteretic regime of the HPART model.  The high overlap 
hints at the possibility that the controlling factor of the HPART model could effectively be the mechanism of the BPART model governing the latter's buffer regime. This hint is consistent with the test results based on separate families of hypotheses. 

\begin{figure}[!htbp]
  \centering
  \includegraphics[width=1\linewidth]{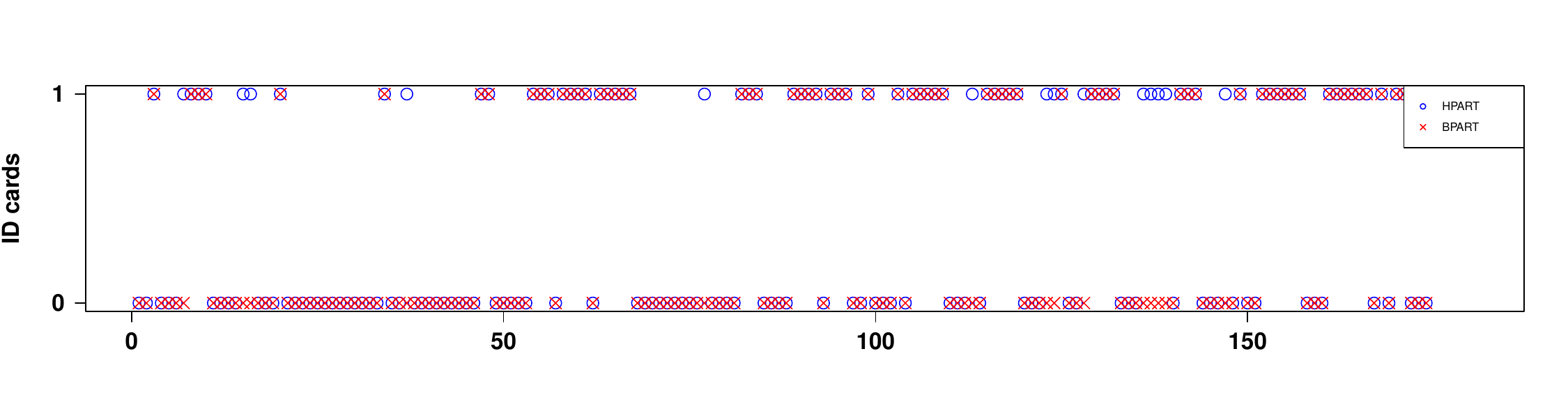}
  \caption{The sequence of datum's ID cards}
  \label{ID cards1}
\end{figure}

To probe further, we plot the binary sequence of datum's ID cards generated by the two models separately (i.e. $R_t$) in Fig.~\ref{ID cards1}.
Table \ref{Contingency Table1} presents the contingency table of ID cards derived from the two models.
Using the Bayesian Information Criterion (BIC) \citep{Katz1981}, the ID sequences of both models exhibit first-order Markov chain properties. Furthermore, a likelihood ratio test (\cite{Anderson1957}, \cite{Billingsley1961}) was conducted to examine the null hypothesis of the two binary sequences being derived from the same Markov chain. The test yielded a test statistic of 2.46 and a $p$-value of 0.29, leading to non-rejection of the null hypothesis. 
Therefore, within the confines of our data, the two models may be considered equivalent in terms of their buffer and hysteretic mechanisms.



\begin{table}[!htbp]
  \centering
  \caption{Contingency Table of ID cards}
  \begin{tabular}{lccc}
    \toprule
    \quad  &BPART:ID=1 &BPART:ID=0 &total\\
    \midrule
    HPART:ID=1 &72 &14 &86\\
    HPART:ID=0 &0 &94 &94\\
    total &72 &108 &180\\
    \bottomrule
  \end{tabular}
  \label{Contingency Table1}
\end{table}



\FloatBarrier
\subsection{Hepatitis B Data}
\begin{figure}[!htbp]
  \centering
  \includegraphics[width=0.9\linewidth]{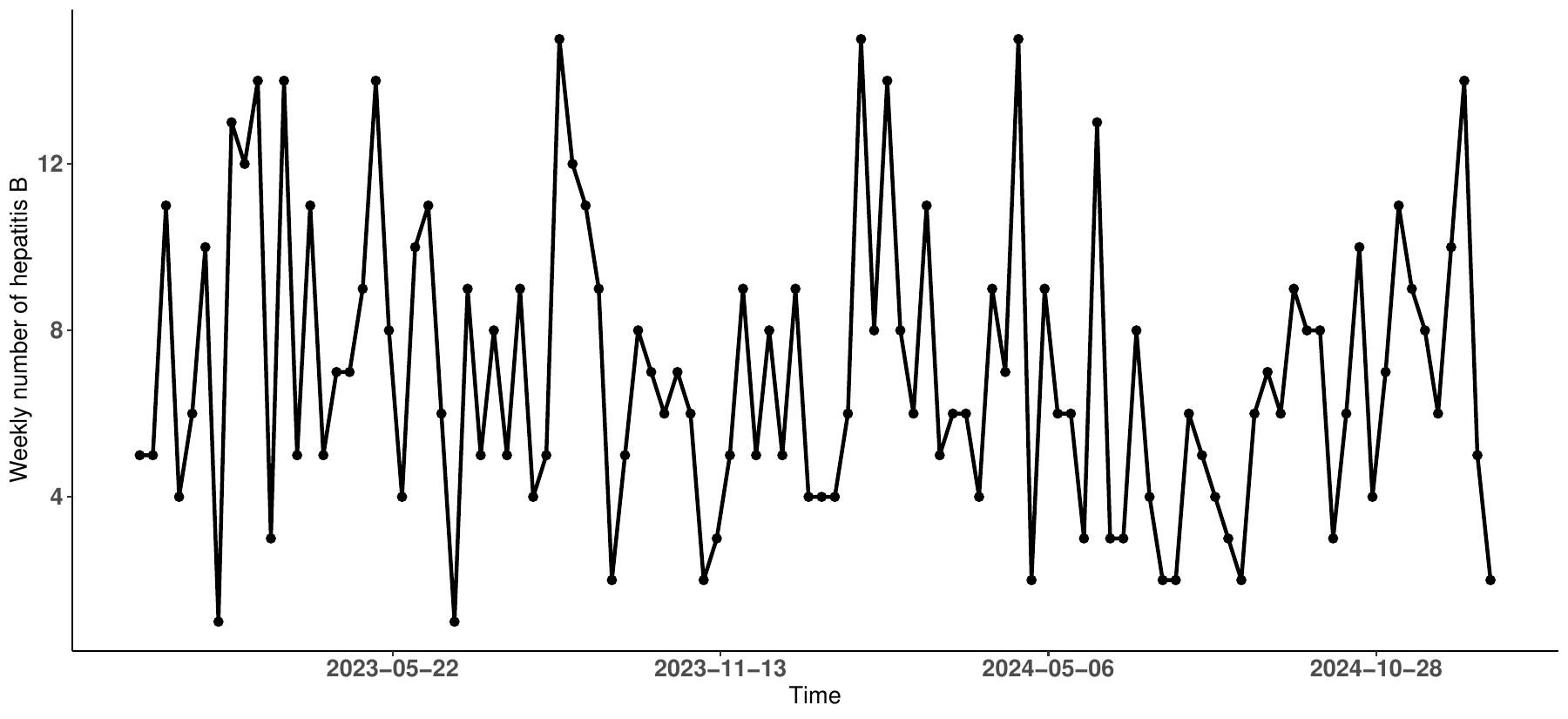}
  \caption{The series of the weekly number of hepatitis B during
2023-2024.}
  \label{hepatitis}
\end{figure}

Consider 
the weekly number of hepatitis B cases reported in
the state of Bremen from January 2023 to February 2024. We shall not repeat reasons for fitting the various models. Now, the data consists of 104 observations, available at  https://survstat.rki.de. The data is plotted in Fig.~\ref{hepatitis}.
We use the first 94 data to fit models and the last 10 data for rolling out-of-sample one-step-ahead predictions. 
The fitted PAR model is 
\begin{flalign*}
&\hspace{15mm}\mathcal{L}(y_t\mid\mathcal{F}_{t-1})=\mathrm{Poisson}(\lambda_t),\\
&\lambda_t = 0.02_{(0.38)} + 0.001_{(0.04)} y_{t-1}+0.99_{(0.09)}\lambda_{t-1},
\end{flalign*}
The fitted SETPAR model is
\begin{flalign*}
&\hspace{15mm}\mathcal{L}(y_t\mid\mathcal{F}_{t-1})=\mathrm{Poisson}(\lambda_t),\\
&\lambda_t =
\begin{cases}
\begin{aligned}
&0.01 && + 0.003 y_{t-1} && + 0.92 \lambda_{t-1} \\[-1ex]
& (5.71) &&\quad (0.43) &&\quad (0.84)
\end{aligned} 
&\mathrm{if} \hspace{2mm} y_{t-1} \leq 4, \\[20pt]
\begin{aligned}
&3.63 && + 0.05 y_{t-1} && + 0.44 \lambda_{t-1} \\[-1ex]
& (3.76) &&\quad (0.09) &&\quad (0.54)
\end{aligned} 
&\mathrm{if} \hspace{2mm} y_{t-1} > 4.
\end{cases}
\end{flalign*}


Now, consider tests between the BPART model and the HPART model.
\begin{flalign}\label{BPART2}
H_0 : \quad\notag&\mathcal{L}(y_t\mid\mathcal{F}_{t-1})=\mathrm{Poisson}(\lambda_t),\\
&\lambda_t =
\begin{cases}
\begin{aligned}
&0.0001 && + 0.25 y_{t-1} && + 0.88 \lambda_{t-1} \\[-1ex]
& (1.51) &&\quad (0.15) &&\quad (0.23)
\end{aligned} 
&\mathrm{if} \hspace{2mm} R_t = 1, \\[20pt]
\begin{aligned}
&0.001 && + 0.08 y_{t-1} && + 0.85 \lambda_{t-1} \\[-1ex]
& (1.66) &&\quad (0.08) &&\quad (0.25)
\end{aligned} 
&\mathrm{if} \hspace{2mm} R_t = 0.
\end{cases}\qquad
R_{t} =
	\begin{cases}
		1, & \mathrm{if} \hspace{2mm} y_{t-1}\leq 5,\\
		0, & \mathrm{if} \hspace{2mm}  y_{t-1}> 7,\\
		R_{t-1}, & \mathrm{otherwise},
	\end{cases}
\end{flalign}

\begin{flalign}\label{HPART2}
\widetilde{H}_0 : \quad\notag&\mathcal{L}(y_t\mid\mathcal{F}_{t-1})=\mathrm{Poisson}(\lambda_t),\\
&\lambda_t =
\begin{cases}
\begin{aligned}
&0.001 && + 0.24 y_{t-1} && + 0.92 \lambda_{t-1} \\[-1ex]
& (1.76) &&\quad (0.19) &&\quad (0.25)
\end{aligned} 
&\mathrm{if} \hspace{2mm} \{\Delta y_{t-1} \geq 2, y_{t-1} \leq 7\} \hspace{2mm} \mathrm{or} \hspace{2mm} \{\Delta y_{t-1} < 2, y_{t-1} \leq 5\}, \\[20pt]
\begin{aligned}
&0.001 && + 0.002 y_{t-1} && + 0.95 \lambda_{t-1} \\[-1ex]
& (1.54) &&\quad (0.06) &&\quad (0.23)
\end{aligned} 
&\mathrm{if} \hspace{2mm} \{\Delta y_{t-1} \geq 2, y_{t-1} > 7\} \hspace{2mm} \mathrm{or} \hspace{2mm}\{\Delta y_{t-1} < 2, y_{t-1} > 5\}.
\end{cases}
\end{flalign}

\begin{table}[!htbp]
  \centering
  \caption{Test results}
  \begin{tabular}{lrrr}
    \toprule
    \quad  &$\alpha = 0.1$ &$\alpha = 0.05$ &$\alpha = 0.01$ \\
    \midrule
    Test (\ref{BPART2}) &Not rejected &Not rejected &Not rejected \\
    Test (\ref{HPART2}) &Not rejected &Not rejected &Not rejected \\
    \bottomrule
  \end{tabular}
  \label{test result2}
\end{table}

 Table \ref{test result2} shows that we reach similar conclusions for the tests as in the case of the escape custody data.

Before moving forward, an initial inspection of the four models shows the following features: (i)  except for  the SETPAR model, the coefficient for $\lambda_{t-1}$ is largely comparable across the other three models, implying that the differences in their $\lambda_{t}$ are primarily attributable to the combined influence of the intercept and $y_{t-1}$; 
(ii) the BPART model and the HPART model share the same $\{r,s\}$ parameters and a similar linear dynamics in the lower regime, but a different linear dynamics in the upper regime, a feature that mighy have an impact  on its predication performance; (iii) the PAR model practically stands out on its own, and  we might arguably consider $\lambda_t \approx \lambda_{t-1}$, in which case the PAR model might perhaps represent a steady state; (iv) the two linear dynamics of the SETPAR model bear no resemblance to those of either the BPART model or the HPART model; (v) in both the fitted BPART model and the fitted HPART model, 
the intercept parameters $\omega_1,\omega_2$ are very close to 0, and, for the upper regime of the HPART model,  
the coefficient of $y_{t-1}$  is also  close to 0 implying that $\lambda_t \approx \lambda_{t-1}$ possibly leading to a different prediction performance compared with the BPART model;    
(vi) similar back-of-envelope calculations give the following approximate combined effects: PAR model (0.03), SEPTAR model (0.03, 4.0), BPART model (1.8, 0.5) and HPART model (1.7, 0.01).

As for the performance of prediction, we get the following results.
\begin{table}[!htbp]
  \centering
  \caption{Summary of prediction error}
  \begin{tabular}{lrrrr}
    \toprule
    \quad  &PAR &SETPAR &BPART &HPART  \\
    \midrule
    MSE &13.64 &11.89 &11.89 &10.81\\
    MAE &2.97 &2.87 &2.66 &2.53\\
    \bottomrule
  \end{tabular}
  \label{...}
\end{table}

The PAR model performs the worst among the four models in the prediction exercise. This might be due to feature (iii) mentioned above. Despite feature (iv), the prediction performance of the SETPAR model is, somewhat to our surprise, identical to that of the BPART model. 
Overall, the HPART model performs by far the best in the prediction exercise, substantially improving the PAR model by 20.7\% and 14.8\% in terms of MSE and MAE respectively. The fact that the BPART model and the HPART model are different seems to underline their difference in prediction performance. 

The coefficient of $y_{t-1}$ captures the short-term feedback effect of last week’s realized cases: In the BPART model, this effect is stronger($\alpha_1=0.25$) in the lower regime, suggesting that even small counts may propagate into the next week. In contrast, the effect is weaker ($\alpha_2=0.08$) in the upper regime, indicating that the process is less sensitive to additional cases once the epidemic is already elevated, possibly reflecting saturation or control measures. A high coefficient of $\lambda_{t-1}$ indicates persistence of the potential outbreak intensity from the previous week. This implies that the number of hepatitis B cases has significant inertia, and last week's risk level impacts this week's risk level.

For HPART model, $\alpha_1=0.24$ and $\alpha_2=0.002$ implying that once the number of hepatitis B cases is high, the consitional mean of the process of next week is driven almost entirely by its previous intensity level. This means that the subsequent direct impact of the high number of hepatitis B cases is weakened, like cases on August 19, 2023 and January 17, 2024. Under high incidence condition, $\beta_2=0.95$ is larger, indicating that once a high incidence state is entered, the risk declines very slowly, showing a significant hysteresis effect, and it takes a long time to return to a low level. In addition, the relatively small magnitude of $c$ and the closeness of the two thresholds, namely $r=5$, $s=7$,  suggest that the model will switch to a high state more frequently and predict longer high-risk periods. In practical public health contexts, this setting is equivalent to a `precautionary strategy': as long as there is a clear upward trend in cases, the alert level is raised, rather than waiting for a larger increase before switching.


Finally, among the first 94 data, 20 are in the buffer/hysteretic area and 14 of them have the same ID card 
$1/0$ marking.  This high proportion at 70\% hints at the possibility that, for the hepatitis B data,  the controlling factor of the HPART model could be effectively the mechanism of the BPART model governing the latter's buffer regime. This hint is consistent with the test results based on separate families of hypotheses. However, despite 
the possibly similar common controlling factor, unlike the previous case, for the present case feature (v) noted above appears to have allowed the HPART model to edge over the BPART model in prediction performance. 

\begin{figure}[!htbp]
  \centering
  \includegraphics[width=1\linewidth]{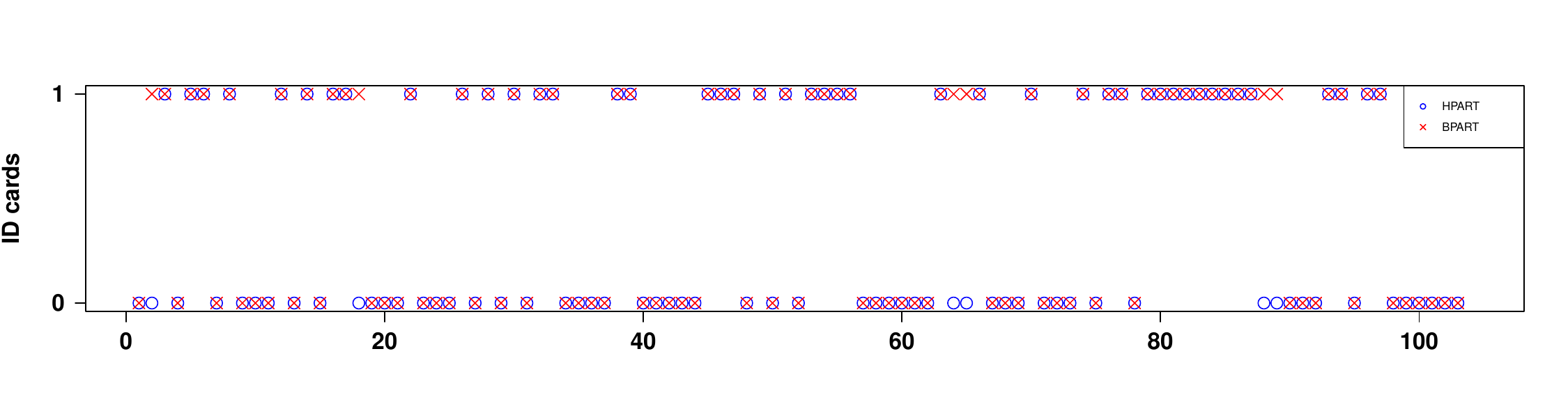}
  \caption{The sequence of datum's ID cards}
  \label{ID cards2}
\end{figure}

The binary sequence of datum's ID cards generated separately by the HPART and BPART models are plotted in Fig.~\ref{ID cards2}.
Table \ref{Contingency Table2} presents the two models’ ID card contingency table.
Using the BIC, both sequences are identified to have a Markov chain order of 0, which implies they can be treated as independent data. The $p$-value of the exact binomial test is 0.03. At the significance level of 0.05, we reject the null hypothesis that the two ID card sequences are identical. This suggests that within the confines of our data, the two models differ with respect to their buffering and hysteretic mechanisms.

\begin{table}[!htbp]
  \centering
  \caption{Contingency Table of ID cards}
  \begin{tabular}{lccc}
    \toprule
    \quad  &BPART:ID=1 &BPART:ID=0 &total\\
    \midrule
    HPART:ID=1 &45 &0 &45\\
    HPART:ID=0 &6 &53 &59\\
    total &51 &53 &104\\
    \bottomrule
  \end{tabular}
  \label{Contingency Table2}
\end{table}

\FloatBarrier
\section{Conclusion and Discussion}\label{conclusion}
In this article, we have proposed a new hysteretic Poisson autoregressive model with thresholds and further analysed the existing buffered Poisson autoregressive model with thresholds. We have
studied the properties of the maximum likelihood estimators of parameters of both models in a unified manner. Most importantly, we have demonstrated the advantages of incorporating a hysteretic regime in the model by (i) providing a physically meaningful mechanism governing the regime-switching therein, (ii) a plausible interpretation and (iii) a better
prediction performance. 

Indeed, we have showcased two real-data studies to confirm the importance and effectiveness of a hysteretic regime. Further, these studies have also provided evidence in support of the view that the hysteretic mechanism of the HPART model could, in some cases, reveal the 
origin of the switching mechanism operated by the BPART model in the buffer regime. Specifically, we have highlighted instances when the $R_{t-1}$ of the BPART model is underlined by the controlling factor $I(\Delta y_{t-1} < c)$ of the HPART model.  


The figure below gives a graphic representation of the hysteretic mechanism of the HPART model. The pink area represents the lower regime while the blue area the upper regime. The two vertical dashed lines represent the two thresholds $r$ and $s$, while the diagonal dashed lines represent the line of first-order difference equal to $c$, where $c$ is taken as $-2$ for illustration. Similarly, the maximum range of $y$ is set at 15 just for illustration. Note that the hysteretic regime for a HPART model is neatly separated into a pink region and a blue region. In general, for the buffer regime of a BPART model, the pink-blue colour is expected to be more randomly distributed.    

In a HPART model, when $r<y_{t-1}\leq s$, which linear model that $y_t$ will follow is determined by the controlling factor $y_{t-1}-y_{t-2}.$ If a BPART model and a HPART model fitted to the same time series share almost the same thresholds, then a substantial overlap of the ID cards of data in the hysteretic/buffer regime suggests that the two models might have a similar controlling factor. 


\begin{figure}[!htbp]
  \centering
  \includegraphics[width=0.7\linewidth]{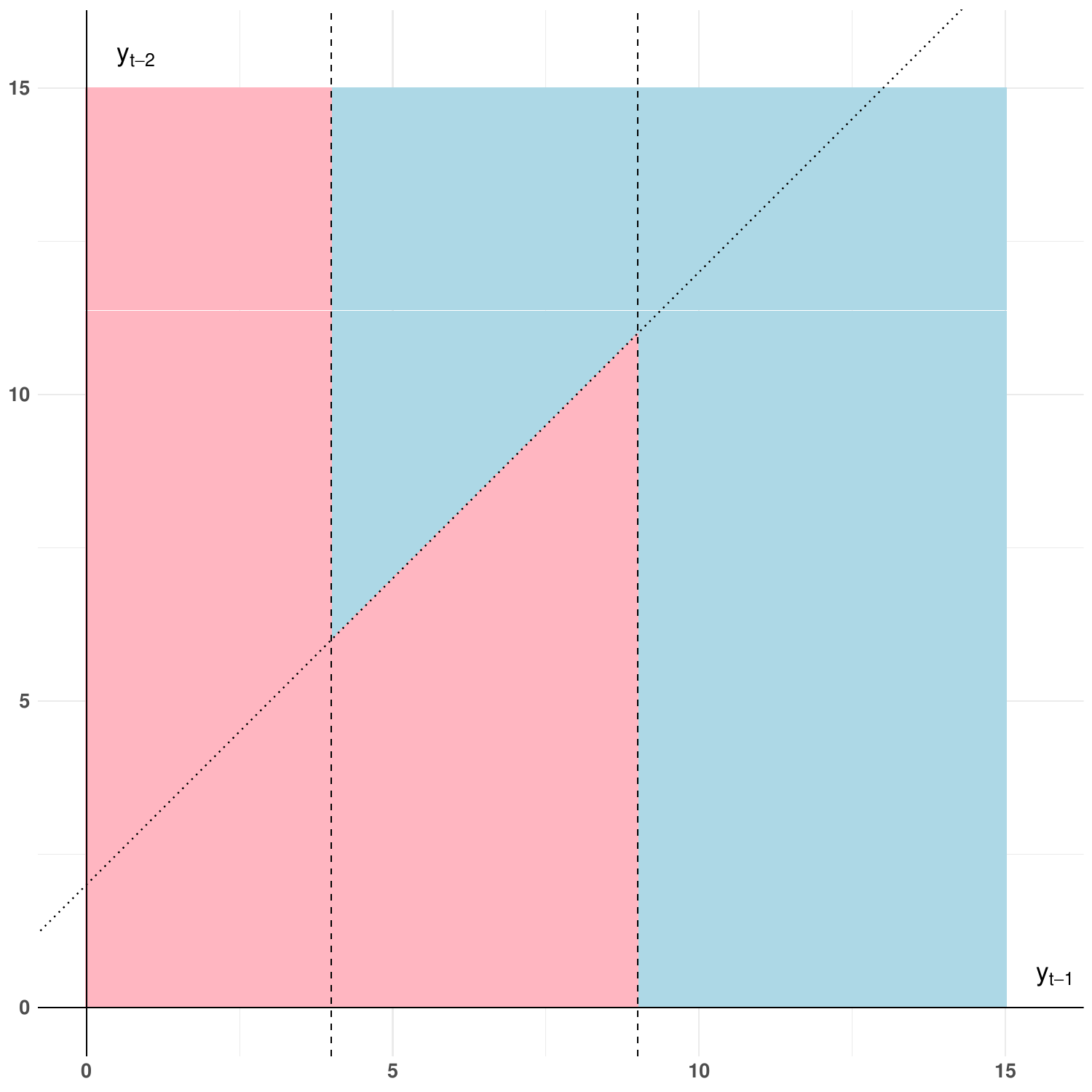}
  \caption{Schematic diagram: hysteretic regime lying between the two vertical dotted lines.}
  \label{Schematic diagram}
\end{figure}

\section*{Supplementary Material}
The supplementary material contains proofs of all theorems in the paper with some useful technical lemmas. 

\bibliographystyle{chicago}
\bibliography{main}

\end{document}